\documentstyle[12pt]{article}

\begin{document}

\author{Sukratu Barve\thanks{%
e-mail: sukkoo@tifrvax.tifr.res.in} and T. P. Singh\thanks{%
e-mail: tpsingh@tifrvax.tifr.res.in}  \\ 
Theoretical Astrophysics Group\\Tata Institute of Fundamental
Research,\\Homi Bhabha Road, Bombay 400 005, India}
\title{Naked Singularities and the Weyl Curvature Hypothesis}
\date{}
\maketitle

\begin{abstract}
\noindent We examine the growth of the Weyl curvature in two examples of
naked singularity formation in spherical gravitational collapse - dust and
the Vaidya spacetime. We find that the Weyl scalar diverges along outgoing
radial null geodesics as they meet the naked singularity in the past. The 
implications of this result for the Weyl curvature hypothesis are
discussed. We mention the possibility that although classical general
relativity admits naked singularity solutions arising from gravitational
collapse, the second law of thermodynamics could forbid their occurrence
in nature. The method can also be used to compare the relative importance of
initial data and that of the energy-momentum tensor in deciding the metric
solution in any general case.
\end{abstract}

\newpage

\section{Introduction}

As Penrose has emphasized \cite{pen}, a fundamental explanation for the
second law of thermodynamics can be had only after we first understand why
the entropy of the Universe was extremely low in the beginning. There is
good cosmological evidence that in the very early epochs, the matter
distribution itself was in a thermal equilibrium state with maximal entropy.
Hence the matter fields by themselves cannot provide an explanation for the
low initial entropy. The low entropy constraint has to come from the
spacetime geometry - a suitably constructed geometrical quantity plays the
role of gravitational entropy. This gravitational entropy has to be
initially so very small that the net sum of the matter and gravitational
entropy takes a value far less than it could possibly have. As the Universe
evolves, the gravitational entropy increases as a result of clumping of
matter and formation of irregularities. The overall entropy of the Universe
also increases.

A plausible geometrical quantity which could play the role of gravitational
entropy is the Weyl curvature tensor \cite{pen}. This is because at the
initial Big Bang singularity, the Ricci curvature necessarily diverges,
whereas the Weyl curvature is zero, if the Universe is assumed to be
Friedmannian. The development of inhomogeneities during evolution leads to
an increase in the Weyl curvature. Ultimately, it is expected that
inhomogeneities lead to the formation of gravitational singularities. If
cosmic censorship holds, these final singularities will be black-holes, and
as a result of the continuous increase of the Weyl tensor, we can expect
that the Weyl tensor will diverge at the black-hole singularity. Thus the
structure of the final singularity is very different from the initial
singularity, in so far as the behaviour of the Weyl curvature is concerned.

In this way of looking at things, the validity of the second law requires
that at the initial cosmological singularity, the Weyl curvature be zero, or
at least ignorable in comparison with the Ricci curvature. This is the
essence of Penrose's Weyl curvature hypothesis. Various important
investigations have been carried out in order to formulate a precise version
of the hypothesis \cite{new}. For natural reasons, these investigations are
largely concerned with the initial cosmological singularity. Here we are
interested in the Weyl hypothesis from a different viewpoint, namely the
possible relevance of this hypothesis for naked singularities occurring in
gravitational collapse of compact objects. A naked singularity should be
regarded as a `T.I.F.' as well as a `T.I.P.', because both ingoing and
outgoing geodesics terminate at such a singularity. In one particular
version of the hypothesis \cite{pen2}, Penrose enquires if given a naked
singularity, the Weyl curvature would diverge along ingoing geodesics, and
go to zero along outgoing geodesics terminating at the singularity. In other
words, one is asking if the Weyl hypothesis that is usually applied to an
initial cosmological singularity would also hold for a naked singularity
forming in collapse. It is this version of the hypothesis that we examine in
our brief note. Considering the similarity between a naked singularity and
an initial cosmological singularity in some, though certainly not all,
respects this version of the hypothesis deserves further examination.

This issue becomes interesting in the light of examples of naked
singularities that have been found in recent years. Here we analyse the
evolution of the Weyl tensor in the Tolman-Bondi spacetime, and in the
Vaidya spacetime, both of which admit naked singularities for certain
initial data. As is known, in both cases, it is only the {\it central}
shell-focusing singularity which can be naked. For definiteness, we assume
that it is the Weyl scalar $C\equiv C^{abcd}C_{abcd}$ which has to be zero
at the initial singularity. We find that the Weyl scalar remains zero at the
center for all epochs prior to the formation of the singularity. (The
singularity is defined by the divergence of the Kretschmann scalar). Exactly 
at the (naked) singularity, however, the scalar blows up, both along ingoing 
as well as outgoing null geodesics. This appears to suggest that the version 
of the Weyl hypothesis under consideration does not hold if such naked 
singularities occur.

A knowledge of the evolution of Weyl curvature is of interest also for
reasons other than the Weyl hypothesis. Various criteria for judging the
seriousness of a naked singularity have been proposed - these include
curvature strength and stability of the Cauchy horizon, among others. In
addition to these, the divergence of the Weyl curvature at the naked
singularity could imply that the singularity is to be taken seriously,
because the Weyl tensor is not related to the local matter distribution, and is
more directly associated with that part of the gravitational field which is
determined by the matter non-locally. On the other hand, if the Weyl scalar
were to be zero at the naked singularity, one could possibly conclude that 
these singularities are associated with the matter distribution.

The paper is organised as follows. In Section II we study the evolution of
the Weyl tensor in the Tolman-Bondi dust collapse. We also use this
opportunity to present a much simplified demonstration of the occurrence of
the naked singularity. From this demonstration one can see that while the
early pioneering works on inhomogeneous dust collapse established the
formation of a naked singularity through difficult calculations, we now have
elementary ways of rederiving those early results. In Section III the Weyl
calculation is repeated for the Vaidya spacetime. The possible implications
of these results are discussed in Section IV.

\section{Weyl tensor for dust collapse}

The Tolman-Bondi metric for dust collapse in comoving coordinates $%
(t,r,\theta ,\phi )$ is

\begin{equation}
\label{met}ds^2=dt^2-\frac{R^{\prime 2}}{1+f(r)}dr^2-R^2(r,t)d\Omega ^2 
\end{equation}
where $R(t,r)$ is the area radius, and the free-function $f(r)$ labels
bound, marginally bound and unbound models, depending on whether $f(r)$ is
negative, zero or positive, respectively. The field equations are 
\begin{equation}
\label{eqf}\epsilon (r,t)=\frac{F^{\prime }(r)}{R^2(r,t)R^{\prime }(r,t)}%
,\qquad \dot R^2=f(r)+\frac{F(r)}R, 
\end{equation}
where dot and prime indicate partial derivative with respect to $t$ and $r$,
respectively. The quantity $F(r)$ is equal to two times the mass inside the
sphere labelled $r$. The metric solution in terms of the parameters above
yields a singularity in the spacetime when $R(r,t)=0$, where the Kretschmann
scalar diverges.

As has been shown earlier \cite{dj}, the central shell-focusing singularity
is naked if the radial null-geodesic equation

\begin{equation}
\label{roo}X=\lim \limits_{r\rightarrow 0,R\rightarrow 0}\frac{R^{\prime
}(X,r)}{\alpha r^{\alpha -1}}\left\{ 1-\sqrt{\frac{\Lambda _0}X}\right\} 
\end{equation}
admits one or more positive real roots $X_0$. This equation is written by
first eliminating the variable $t$ in favour of $R$, and then eliminating $R$
in favour of $X\equiv R/r^\alpha $, where $X$ is the tangent to a possible
outgoing geodesic. The constant $\alpha >1$ is chosen so that $R^{\prime
}/r^{\alpha -1}$ has a unique, finite limit as $R\rightarrow 0,r\rightarrow
0 $. The constant $\Lambda _0$ is the limiting value of $\Lambda
(r)=F(r)/r^\alpha $ as $r\rightarrow 0$.

At this stage the calculations in earlier papers \cite{dj}, \cite{sin}
proceed by giving a detailed expression for $R^{\prime }$ - this is the most
involved part of the calculation. The simplification we present here is in
the calculation of $R^{\prime }$. Consider first the marginally bound case,
for which the solution is

\begin{equation}
\label{mar}R^{3/2}=r^{3/2}-\frac 32\sqrt{F(r)}t. 
\end{equation}
An initial scaling $R=r$ at the starting epoch $t=0$ of the collapse has
been assumed. From this equation, evaluate $R^{\prime }$ and substitute for $%
t$ from ($\ref{mar}$)$.$ In the resulting expression, substitute $%
R=Xr^\alpha $, and perform a Taylor expansion of $F(r)$ around $r=0$, so as
to retain the leading non-vanishing term. We get that near $r=0$,

\begin{equation}
\label{rpr}\frac{R^{\prime }}{r^{\alpha -1}}=X+\frac \theta {\sqrt{X}%
}r^{q+3/2-3\alpha /2}. 
\end{equation}
Here, $\theta =-qF_q/3F_0$, and $q$ is defined such that in a series
expansion of the initial density $\rho (r)$ near the center, the first
non-vanishing derivative is the $q$th one (=$\rho _q$), and $F_q=\rho
_q/(q+3)q!$. Now a unique definition of $\alpha $ is obtained by setting the
power of $r$ as zero in the second term, giving $\alpha =1+2q/3$. This then
reproduces the result of the $R^{\prime }$ calculation performed earlier 
\cite{sin}, in a simpler manner. From this point on, the naked singularity
analysis through evaluation of roots of (\ref{roo}) proceeds as before, but
the overall calculation is now more economic.

For the non-marginally bound case, the solution of the Tolman-Bondi
equations is

\begin{equation}
\label{nmar}R^{3/2}G(-fR/F)=r^{3/2}G(-fr/F)-\sqrt{F(r)}t 
\end{equation}
where $G$ is a known elementary function \cite{dj}. $R^{\prime }$ can be
evaluated as before, and then we eliminate $t$ and substitute $R=Xr^\alpha $%
. The power-series expansion for $f(r)$ near $r=0$ is of the form \cite{sin}

\begin{equation}
\label{fx}f(r)=f_2r^2+f_3r^3+f_4r^4+... 
\end{equation}
which implies that the argument ($-fR/F$) of $G$ on the left-side of ($\ref
{nmar})$ goes to zero as $r\rightarrow 0$. The derivative of $G(-p)$
w.r.t. its argument $p\equiv rf/F$ is obtained by differentiating (\ref{nmar}%
), which gives

$$
\frac{dG(-p)}{d(-p)}=\frac{3G}{2p}-\frac 1{p\sqrt{1+p}}. 
$$
Using this, one can now perform a Taylor-expansion of $G,F$ and $f$ about $%
r=0$ to get exactly the same expression for $R^{\prime }$ as given in (\ref
{rpr}) above, where $\theta $ is now given by

\begin{equation}
\label{the}\theta =q\left( 1-\frac{f_2}{2F_0}\right) \left(
G(-f_2/F_0)\left[ \frac{F_q}{F_0}-\frac{3f_{q+2}}{2f_2}\right] \left[ 1+%
\frac{f_2}{2F_0}\right] +\frac{f_{q+2}}{f_2}-\frac{F_q}{F_0}\right) . 
\end{equation}
The constant $q$ is now defined such that the first non-vanishing derivative
of the initial density is the $q$th one, and/or the first non-vanishing term
in the expansion for $f(r)$, beyond the quadratic term, is of order $r^{q+2}$%
. The constant $\alpha $ is again equal to $(1+2q/3)$. Thus the $R^{\prime }$
calculation is simplified for the non-marginal case as well.

We return now to the calculation of the Weyl tensor, which is defined as 
$$
C_{abcd}=R_{abcd}+ g_{a[d}R_{c]b}+g_{b[c}R_{d]a} +\frac
13Rg_{a[c}g_{d]b}, 
$$
where square brackets denote antisymmetrization. The nonvanishing independent 
components for the Tolman-Bondi metric are

\begin{equation}
\label{wey}
\begin{array}[t]{l}
C_{1010}= 
\frac{R^{\prime 2}}{1+f}\left( \frac F{R^3}-\frac{F^{\prime }}{3R^2R^{\prime
}}\right) \\ C_{2020}=-\frac F{2R}+ 
\frac{F^{\prime }}{6R^{\prime }} \\ C_{2121}=- 
\frac{F^{\prime }R^{\prime }}{6(1+f)}+\frac{F(r)R^{\prime 2}}{2R(1+f)} \\ 
C_{3030}=C_{2020}\sin ^2\theta \\ 
C_{3131}=C_{2121}\sin ^2\theta \\ 
C_{3232}=\left( -FR+\frac{F^{\prime }R^2}{3R^{\prime }}\right) \sin ^2\theta 
\end{array}
\end{equation}
The Weyl scalar is given by

\begin{equation}
\label{sca}C=\frac{12}{R^4}\left( \frac FR-\frac{F^{\prime }}{3R^{\prime }}%
\right) ^2. 
\end{equation}
At the start of evolution at $t=0$, where the scaling $R=r$ has been chosen,
we see that the Weyl scalar at $r=0$ is zero (the mass-function $F(r)$ grows
as $r^3$ near $r=0$.) Using the above Tolman-Bondi solution we find that the
scalar remains zero at $r=0$ throughout the non-singular phase of the
evolution. However, at the epoch of naked singularity formation, we know
that in the neighborhood of the singularity, we have along an outgoing
geodesic the relation $R=X_0r^\alpha $, $X_0$ being the finite tangent.
Also, $R^{\prime }$ is given by (\ref{rpr}), with $X=X_0$. With these
substitutions, we see that in the limit of approach to the singularity, the
scalar is equal to

\begin{equation}
\label{wey2}C(X_{0,}r)=\frac{12F_0^2\theta _0^2}{X_0^7\left( X_0+\theta _0/%
\sqrt{X_0}\right) ^2}r^{6(1-\alpha )}. 
\end{equation}
For those initial data that lead to a naked singularity, the range of $%
\alpha $ is $5/3\leq \alpha \leq 3$. Hence we see by letting $r$ go to zero
that the Weyl scalar diverges at the naked singularity along outgoing
geodesics. In so far as ingoing geodesics are concerned, these exist for all
initial data, i.e. $\alpha $ takes all values greater than unity. Hence the
Weyl scalar necessarily diverges along ingoing geodesics too.

The effect of the energy-momentum tensor on the geometry can be found by
evaluating the scalar $R_{\alpha \beta }R^{\alpha \beta }$ and comparing it
to the Weyl scalar. We get in the limit of approach to the naked singularity,

\begin{equation}
\label{ric}R_{\alpha \beta }R^{\alpha \beta }=\frac{F^{\prime 2}}{%
R^{^{\prime }2}R^4}=\frac{9F_0^2}{X_0^4\left( X_0+\theta _0/\sqrt{X_0}%
\right) ^2}r^{6(1-\alpha )}. 
\end{equation}
We find that this scalar diverges at the same rate as the Weyl scalar, and
the latter does not dominate over the former. Notice however that unlike the
Weyl scalar, the invariant $R_{\alpha \beta }R^{\alpha \beta }$ is non-zero
at the initial epoch, and evolves with time. Finally we note that for the
points $r\neq 0$ the Weyl scalar is in general non-zero initially - it then
evolves with time and diverges at the singularity $R(t,r)=0.$ We recall that
the singularity at $r\neq 0$ is necessarily covered.
\newpage
\section{Weyl tensor for Vaidya spacetime}

The collapse of null dust is described by the Vaidya metric

\begin{equation}
\label{vai}ds^2=\left( 1-\frac{2m(v)}r\right) dv^2-2dvdr-r^2d\Omega ^2 
\end{equation}
where $v$ is the advanced time, and $m(v)$ is the mass-function. The
energy-momentum tensor is

\begin{equation}
\label{en}T_{ik}=\epsilon u_iu_k=\frac 1{4\pi r^2}\frac{dm}{dv}u_iu_k 
\end{equation}
where $u_i=-\delta _i^v$ - this represents the radial inflow of radiation
into an initially flat region of spacetime. The spacetime outside the
radiation region is Schwarzschild. In the self-similar model, which is the
case that we consider here, the mass-function in the cloud is given by $%
2m(v)=\lambda v$, i.e. it is linear. A curvature singularity forms when a
shell hits the origin $r=0$. It is known that the singularity at $r=0,v=0$
is naked for $\lambda \leq 1/8$ \cite{vn}.

The non-zero independent components of the Weyl tensor are given by

\begin{equation}
\label{weyv}
\begin{array}{l}
C_{1010}= 
\frac{2m(v)}{r^3} \\ C_{2020}=- 
\frac{m(v)}r\left( 1-\frac{2m(v)}r\right) \\ C_{2120}= 
\frac{m(v)}r \\ C_{3232}=-2m(v)r\sin ^2\theta \\ 
C_{3030}=C_{2020}\sin ^2\theta \\ 
C_{3130}=C_{2120}\sin ^2\theta 
\end{array}
\end{equation}
The Weyl scalar is given by

\begin{equation}
\label{weys}C(v,r)=\frac{48m^2}{r^6}=\frac{12\lambda ^2v^2}{r^6}. 
\end{equation}
Prior to the formation of the singularity, the scalar is zero at the inner
edge $v=0$, $r\neq 0$ of the cloud. Naturally, it is also zero in the flat
region to the interior of the collapsing cloud. When the naked singularity
forms, the finite tangent along an outgoing geodesic from the singularity is
defined as $X=v/r$ (see \cite{djv}). Thus we get

\begin{equation}
\label{weysv}C(X_0,r)=\frac{12\lambda ^2X_0^2}{r^4} 
\end{equation}
where $X_0$ is the limiting value of the tangent. The Weyl scalar diverges
at the naked singularity. The scalar $R_{\alpha \beta }R^{\alpha \beta }$ is
however zero, and in this case, the Weyl scalar dominates.

\section{Discussion}

Our aims in this paper were twofold. Firstly, to compute the Weyl tensor in
the approach to the naked singularity in the Tolman-Bondi and Vaidya
spacetimes, and secondly to consider what the results could mean for the
Weyl curvature hypothesis. We find that in these models, the Weyl scalar
behaves in a peculiar manner, remaining zero at the origin throughout the
non-singular phase of the evolution, and then suddenly jumping to infinity
along outgoing geodesics, at the epoch of singularity formation. The same is
true along ingoing geodesics. In so far as the Weyl hypothesis is concerned,
it appears reasonable that the hypothesis apply in the same manner to
cosmological initial singularities and to naked singularities \cite{pen2}.
In the examples considered here, we find that the curvature structure of the
T.I.F.s is similar to that of the T.I.P.s, which does not support the
hypothesis. Also, since naked singularities are expected to arise in
strongly inhomogeneous situations, for which the Weyl tensor is typically
high, one could speculate that the Weyl scalar should generically diverge
in the past along outgoing geodesics, rather than go to zero. 

If we do assume that the Weyl hypothesis must be valid, then
these results offer an interesting possibility, namely that the second
law of thermodynamics could forbid the occurrence of naked singularities in
nature. If the Weyl hypothesis applies both to cosmological initial
singularities and to naked singularities arising from gravitational collapse,
then validity of the second law requires that only those naked singularities
can actually occur in nature for which the Weyl scalar goes to zero along
outgoing geodesics. The second law would hence disallow those naked singular
solutions which disagree with the Weyl hypothesis. If the Weyl scalar 
diverges generically along outgoing geodesics, the second law would 
generically forbid naked singularities from occurring in nature. 

Thus while classical general
relativity admits naked singularities in gravitational collapse, their
occurrence in the real world could be thermodynamically forbidden. 
The situation will then be analogous to the advanced wave solutions in
classical electrodynamics - solutions that are permitted by the theory but
thermodynamically disallowed in the real world. The cosmic censorship
hypothesis would then hold, though not in classical general relativity as
such, but in the real universe, through the intervention of thermodynamics.
However, this would mean attributing a fundamental status to the Weyl Curvature
Hypothesis, which could possibly involve quantum gravity.

Although it was the notion of connection of the Weyl tensor with
non-uniformity of energy momentum distribution which led Penrose to think of
its relation to entropy, it must be kept in mind that the Weyl tensor has a
certain part which is determined by the `initial' conditions of the field
equations. These are the metric and its derivatives specified on a chosen 
space-like surface. The Weyl scalar going to zero from TIF would mean that one
would have to have unique initial conditions on a partial Cauchy surface 
chosen near the singularity formation event, as it approaches the ideal point.
Thus, 
Penrose seems to indicate that there was only a unique choice of the metric 
solution \cite{pen2}, presumably at a limiting space-like surface near the 
ideal point.

Also, given the fact that the divergence of the Weyl scalar and $%
R_{\alpha\beta}R^{\alpha\beta}$ are comparable, if not the Weyl contribution
dominating, one would be led to think that whenever a naked singularity 
occurs, the role played by the initial conditions in deciding the metric of 
the universe is as important as that of the energy momentum tensor, if not 
more.

It must be noted that the conjecture includes all singularities generically and
not just the cosmological ones. We find that it is not true for Tolman Bondi 
dust and Vaidya radiation collapse scenarios. 

This leaves the issue of the factors deciding the initial conditions open.

\end{document}